# Origin and Propagation of Spin-orbit Torques in Pt/Co/Cu/NiFe/Capping Multilayers

Yuming Bai,[1] Rulin Tian,[1] Yue Zhang,[1] Tao Wang[1, a)]

1 School of Integrated Circuits, Huazhong University of Science and Technology, Wuhan, 430074, P. R. China

a) Author to whom correspondence should be addressed: ptw@hust.edu.cn

## Abstract

Spin–orbit torque (SOT) enables efficient current-driven control of magnetization, offering a promising pathway toward low-power spintronic devices. However, the origin and propagation of both damping-like (DL) and field-like (FL) SOTs in complex multilayers remain unclear. Here, we investigated NiFe thickness-dependent of SOT efficiencies in Ta/Pt/Co/Cu/$x$NiFe/Cu/Capping multilayers ($x$ = 1~5 nm; Capping = Pt, Al, and $SiO_2$). By employing a spin rotation geometry, the perpendicularly magnetized Pt/Co/Cu stacks serve as a spin source introducing an unconventional spin polarization orthogonal to the Oersted field, eliminating its contribution and enabling unambiguous extraction of SOTs using planar Hall and polar MOKE measurements. To distinguish bulk and interfacial contributions, we introduce a sample-area-normalized moment $m = m_{NiFe} / S$, accounting for thickness-dependent magnetization and eliminating uncertainties arising from nominal thickness scaling and magnetic dead-layers. We find that DL-SOT follows nearly linear $1/m$ scaling, consistent with rapid spin absorption at the Cu/NiFe interface but exhibits finite $\beta_{\text{SOT}}$ when $1/m$ approaches zero in both Pt- and Al-capped samples, indicating additional interfacial spin-current contributions at Cu/Pt and Cu/Al interfaces. In contrast, $SiO_2$-capped samples show negligible interfacial contributions. Furthermore, FL-SOT deviates markedly from $1/m$ scaling and exhibits a significantly longer spin dephasing length (~1.7 nm) compared to DL-SOT, implying extended propagation across NiFe. Comparative capping-layer studies further corroborate this behavior through interface-dependent spin transport. Our findings clarify the origin and distinct propagation characteristics of DL and FL torques, providing guidelines for engineering interfacial spin-orbit functionalities in ultrathin metallic heterostructures.

The interconversion between charge and spin current is central to modern spintronics, enabling current-driven manipulation of magnetization through spin-orbit torque (SOT).[1–6] In ferromagnet (FM)/nonmagnet (NM) heterostructures, SOT primarily arises from two mechanisms: the bulk spin Hall effect (SHE) and the Rashba-Edelstein effect (REE) at interfaces with broken-inversion symmetry.[7–9] Both mechanisms generate two types of torques with different symmetries: the damping-like (DL) and field-like (FL) torques, which play different roles in magnetization switching and spin transportation.[1,8,10–12] However, the microscopic origin and propagation process of these torques remain incompletely understood, particularly for ultrathin ferromagnetic layers with interfacial disorder and abnormal spin absorption complicate conventional interpretations.[13–15] The FL-SOT, in particular, is highly sensitive to interfacial spin transparency and scattering process, leading to complex and often inconsistent experimental observations.[14,16–19] Moreover, in the conventional SOT measurements, the effective field of the FL-SOT ($h_{SH}^{FL}$) is typically collinear with the charge current induced Oersted field, which makes their separation experimentally challenging. As a result, most previous analyses estimate the $h_{SH}^{FL}$ indirectly by subtracting a numerically calculated Oersted field based on parallel resistor model from the total effective field,[14,20] a procedure that introduces significant uncertainty into quantitative SOT extractions. Hence, there is an urgent need for a method to accurately extract the SOT effective fields.

A fundamental limitation arises from the intrinsic symmetry of the SHE, as shown in Fig.1(a), which an in-plane electrical field $E_x$ generates a spin current $J_s$ flowing out of the plane, but the spin polarization $\sigma$ is confined to the in-plane direction perpendicular to that of $E_x$. Owing to this symmetry restriction, an external magnetic field is required to enable magnetization switching.[7,19] To overcome this limitation, recent studies have focused on field-free switching schemes based on additional symmetry breaking, where generating spin currents with out-of-plane polarization serves as a key mechanism.[21–24] Especially, magnetic materials inherently possess such symmetry-breaking ability via their internal exchange fields and spontaneous magnetization, which can generate the unconventional spin polarization through the spin rotation effect (SRE), as illustrated in Fig.1(b).[25,26] Theoretically, the spin current can be described as:

$$j_s = \frac{\hbar}{2e}\theta j_c \times (m \times \sigma), \qquad (1)$$

where $j_c$ is the in-plane charge current density, $j_s$ is the out-of-plane spin-current density, $\theta$ is the charge-to-spin conversion efficiency, $\hbar$ is the reduced Planck constant, and $e$ is the electron charge. Eq. (1) describes the rotation of spin polarization $\sigma$ around the local magnetization $m$, generating unconventional spin polarization that is symmetry-forbidden in nonmagnetic systems. Similar phenomena have also been experimentally verified in antiferromagnets and altermagnets such as $Mn_3Sn$, $Mn_2Au$, and $RuO_2$, highlighting its universality [27–29]

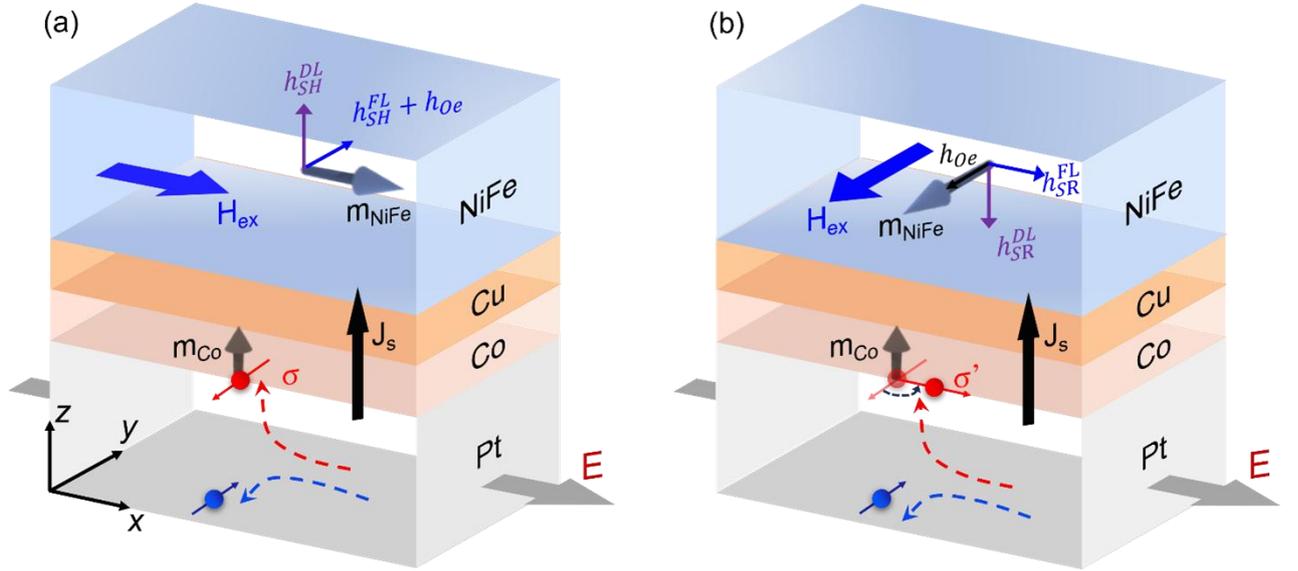

FIG. 1. Schematic of the current-induced SOT effective fields in (a) the conventional spin Hall geometry and (b) the spin Rotation geometry for Pt/Co/Cu/NiFe multilayers. In former case, an electrical field $E_x$ generates an out-of-plane spin current $J_S$ with spin polarization $\sigma = z \times E$, making the FL-SOT effective field $h_{SH}^{FL}$ collinear with the Oersted field $h_{Oe}$. In latter case, the $\sigma$ rotates around Co magnetization $m_{Co}$ and turns into $\sigma' = m_{Co} \times z \times E$, generating an unconventional spin polarization. When the external field $H_{ex} \perp E$, the $h_{SR}^{FL}$ becomes orthogonal to the $h_{Oe}$. Both the DL-SOT effective field $h_{SH}^{DL}$ and $h_{SR}^{DL}$ oriented along the out-of-plane.

In this letter, we utilize an unconventional spin polarization, which originating from the perpendicularly magnetized Pt/Co/Cu layers through SRE to systematically study the origin and propagation process of both FL-SOT and DL-SOT in a typical ferromagnet NiFe.[25,26] In spin Rotation geometry, it excludes Oersted field from torque measurement and enables unambiguous separation of FL and DL torques via planar Hall and polar MOKE measurements. To accurately clarify the bulk and interfacial contributions, we further employ a sample-area ($S$) -normalized moment $m = m_{\mathrm{NiFe}}/S$ that accounts for thickness-dependent magnetization during the SOT vs thickness study, which can get rid of SOT data analysis uncertainty caused by the use of nominal thickness as well as the dead-layer existing at Cu/NiFe interface. Our SOT measurement results reveal that DL-SOT follows nearly linear $1/m$ scaling, consistent with rapid spin absorption at the bottom Cu/NiFe interface but exhibits finite DL-SOT efficiency $\beta_{\mathrm{SOT}}$ when $1/m$ approaches zero in both Pt- and Al-capped samples, indicating additional interfacial spin-current contributions at the Cu/Pt and Cu/Al interfaces. In contrast, $SiO_2$ -capped stacks show negligible interfacial contributions at the Cu/$SiO_2$ interface. Furthermore, FL-SOT deviates markedly from $1/m$ scaling and exhibits a significantly longer spin dephasing length (~1.7 nm) compared to DL-SOT, demonstrating its extended propagation across the NiFe layer. Comparative studies with three individual capping layers further reveal that for NiFe thicknesses below ~1.7 nm, spin currents responsible for FL-SOT reach the top Cu/Capping interface and display different interface-dependent spin transport characteristics, corroborating the extended spin dephasing length of FL-SOT in NiFe. In summary, above results clarify the origin and distinct propagation process of FL and DL torques, providing guidelines for engineering interfacial spin-orbit functionalities in ultrathin metallic heterostructures.

A series of multilayer films with the structure Ta(5)/Pt(5)/Co(0.9)/Cu(4)/ $Ni_{81}Fe_{19}$(x)/Cu(2)/Capping(5) were fabricated, where the numbers in parentheses denote nominal layer thicknesses in nm, $x$ = 1 ~ 5 nm and Capping = Pt, Al, and $SiO_2$. All films were deposited by DC magnetron sputtering in an Ar atmosphere of 2.2 mTorr (base pressure < $7.5 \times 10^{-8}$ Torr. The 4 nm Cu spacer effectively suppresses interlayer exchange coupling between Co and NiFe. The Cu/NiFe interface exhibits negligible magnetic proximity effect,[16,30] allowing the NiFe layer to serve as a clean sensing layer for SOT characterization. A Ta buffer layer promotes perpendicular magnetic anisotropy (PMA) in Co, confirmed by vibrating sample magnetometer (VSM) measurements [Fig.S1(a)], which yield a perpendicular anisotropy field of $H_K$ ~ 0.81 T for the Pt/Co stacks. The films were patterned into Hall bar (500 $\mu$m × 450 $\mu$m) and MOKE (50 $\mu$m × 50 $\mu$m) devices. Contact pads were subsequently capped with Ti(5)/Cu(100)/Au(70) (thickness in nm) using high-vacuum electron-beam-evaporation.

We defined spin Hall geometry and spin Rotation geometry during the SOT measurements, as shown in Fig.1(a) and (b), corresponding to an external magnetic field $H_{ex}$ is parallel and orthogonal to the electric field $E_x$, respectively. In the spin Rotation geometry, the perpendicularly magnetized Pt/Co/Cu layers serve as a spin source, generating an unconventional spin polarization orthogonal to the current-induced Oersted field. This configuration ensures that the resulting FL-SOT effective field ($h_{SR}^{FL}$) lies along $x$-axis, while the Oersted field $h_{Oe}$ is parallel to $m_{NiFe}$ ($y$-axis) which has no contribution to current induced torque. Thus, SOT measurement under spin Rotation geometry allows unambiguous extraction of FL- and DL-SOT components induced by $x$-axis spin polarization $\sigma'$. To accurately distinguish bulk and interfacial contributions, we further introduce a sample-area ($S$) -normalized moment $m = m_{\text{NiFe}}/S$, which properly accounts for thickness-dependent magnetization in the SOT vs thickness analysis and eliminates uncertainties associated with nominal thickness and magnetic dead-layer at the Cu/NiFe interface.

The FL-SOT generated in NiFe layer ($m_{\text{NiFe}}$) was characterized by using the planar Hall effect (PHE) technique, which is particularly sensitive to in-plane effective fields while being nearly immune to out-of-plane contributions owing to the negligible anomalous Hall response of NiFe.[31,32] The representative PHE data is displayed in Fig. 2(b), showing a clear polarity reversal when the magnetization of the Co layer is switched from $m_{Co}$ = +z to $m_{Co}$ = -z, confirming the SRE-induced spin origin. An AC current ($I_{ac}$ = 70 mA, $f$ = 346 Hz) was applied along the Hall bar, and the corresponding second-harmonic Hall voltage ($\Delta V_{2\omega}$) was detected by lock-in amplifier, which can be expressed as:[31]

$$\Delta V_{2\omega} = -\frac{1}{2}\Delta R_P \cos2\varphi I_{ac}\Delta\varphi = -\frac{1}{2}\Delta R_P \cos2\varphi I_{ac}\frac{h_{SR}^{FL}}{|H_{ex}|\pm H_a}sgn[m_x] \quad (2)$$

where $\Delta R_P$ is the change in the planar Hall resistance,[33] $\varphi$ is the in-plane azimuth angle, $H_{ex}$ is the applied external magnetic field, and $H_a$ is the in-plane anisotropy field. The sign function $sgn[m_x]$ represents the direction of the NiFe in-plane magnetization.

As shown in Fig. 2(a), in spin Rotation configuration, the external field $H_{ex}$ is orthogonal to the bias electrical field $E_x$, allowing the elimination of Oersted field interference from SOT measurement. In calibration procedure, a DC current $I_{dc} = 70$ mA was applied to the Hall bar while an AC Oersted field $h_{Cal} = 1.56$ Oe with a frequency of $f = 27.17$ Hz which was generated by a calibration coil. The corresponding first-harmonic Hall voltage ($\Delta V_{1\omega}$) was detected using the same lock-in technique that can be expressed as: [31]

$$\Delta V_{1\omega} = -\Delta R_P cos2\varphi I_{dc}\Delta\varphi = -\Delta R_P cos2\varphi I_{dc} \frac{h_{Cal}}{|H_{ex}|\pm H_a} sgn[m_x] \quad (3)$$

By performing a linear fitting of $\Delta V_{2\omega}$ versus $\Delta V_{1\omega}$ yields a slope which is equal to $\frac{h_{SR}^{FL}}{2h_{Cal}}$, thus, $h_{SR}^{FL}$ is quantitatively determined, as shown in Fig. 2(c).

In general, the bulk and interfacial contributions are commonly distinguished by examining the thickness dependence on the ferromagnetic layer.[8,15] We defined and calculated FL-SOT efficiency $\beta_{SOF} = \frac{h_{SR}^{FL}}{J_c}$, where $J_c$ is the charge current density flowing through all layers except the NiFe layer, obtained via a parallel resistor model, as calculated in Fig. S1(e). As a result, Fig. 2(d) summarizes $\beta_{SOF}$ as a function of the inverse normalized magnetic moment ($1/m$) including samples with Pt, Al, and SiO$_2$ capping layers. The FL-SOT shows pronounced deviations from $1/m$ scaling and a significantly longer spin dephasing length (~1.7 nm) compared to DL-SOT in NiFe, demonstrating its extended propagation across the NiFe layer. The inset shows the extraction of the spin dephasing length $\lambda_{dp}$ of the spin currents responsible for the FL-SOT in NiFe, obtained by fitting $\beta_{SOF}(\lambda_{dp})$ using the spin drift-diffusion model:[8,15,34]

$$\beta_{SOF}(\lambda_{dp}) = Ae^{-d_{eff}/\lambda_{dp}} + C \quad (4)$$

which $A$ denotes the FL-SOT amplitude injected from the bottom Pt/Co/Cu side before spin dephasing in NiFe, and $C$ accounts for thickness-independent contributions arising mainly from the top Cu/Capping interface (Pt, Al, or SiO$_2$).[35,36] The saturation magnetization of NiFe ($M_s$) was estimated by ferromagnetic resonance measurement of a 30nm thick NiFe sample, which was extracted to be $0.99 \pm 0.01$ T [Fig. S3(b)]. The effective thickness $d_{eff}$ is obtained as $d_{eff} = \frac{m}{M_s \cdot S}$, as shown in Fig. S1(b-d). The extracted spin dephasing lengths are $\lambda_{dp}$ = 1.74 ± 0.09 nm (Pt), 1.70 ± 0.13 nm (Al) and 1.73 ± 0.22 nm (SiO$_2$), respectively.

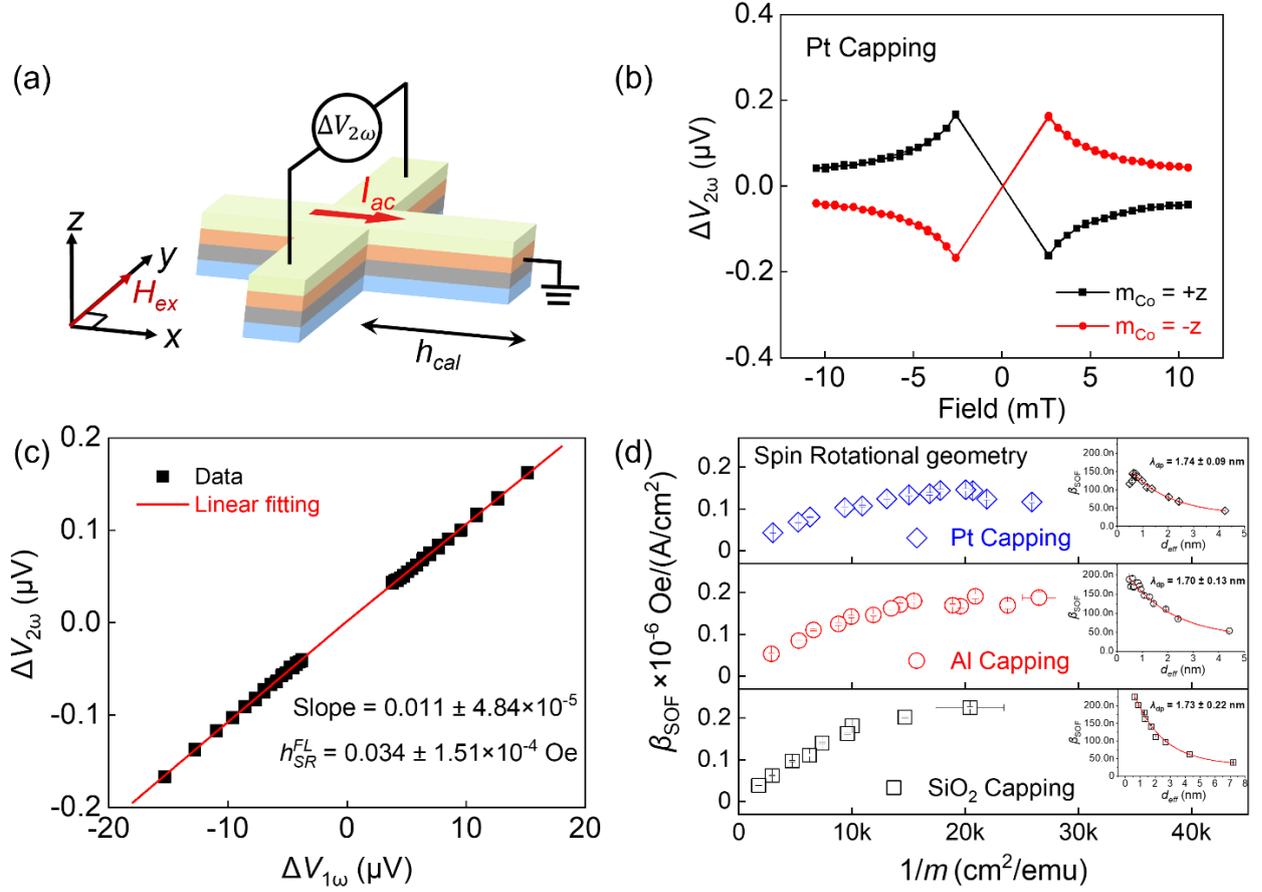

FIG. 2. (a) Schematic of the FL-SOT detection using the planar Hall effect. An AC current $I_{ac}$ and external field $H_{ex}$ are applied orthogonally. A calibration AC field $h_{cal}$ is applied along the *x*-axis. (b) Representative PHE data showing polarity reversal upon switching the Co layer magnetization $m_{Co}$. (c) Linear fitting of $\Delta V_{2\omega}$ versus $\Delta V_{1\omega}$ used to determine the FL-SOT effective field $h_{SR}^{FL}$. (d) FL-SOT efficiency $\beta_{SOF}$ as a function of $1/m$ including samples with Pt, Al and SiO$_2$ capping layers. Inset: Fit of $\beta_{SOF}$ versus the effective NiFe thickness $d_{eff}$, from which the spin dephasing length $\lambda_{dp}$ (~1.7 nm) of the spin currents responsible for the FL-SOT in NiFe is extracted.

More importantly, the three individual capping layer samples yield distinct $\beta_{SOF}$ versus $1/m$ trends, revealing their contrasting spin-transport properties at the Cu/capping interface. For Al-capped samples, $\beta_{SOF}$ deviates from linear $1/m$ scaling and saturates at ultrathin thicknesses. With its weak spin–orbit coupling (SOC) and long spin-diffusion length, Al forms a spin-transparent Cu/Al interface that allows efficient spin current transmission with minimal spin absorption,[36,37] as shown in Fig. 4(a). In contrast, $β_{SOF}$ of Pt-capped samples decreases significantly at ultrathin end. The strong SOC and spin-sink feature of Pt cause substantial spin absorption at both the Cu/Pt interface and within the Pt bulk, thereby it results in a reduction of FL-SOT efficiency [Fig. 4(b)].[38,39] For SiO$_2$-capped samples, $β_{SOF}$ exhibits a gradual attenuation and a less significant deviation from linearity than that of Pt and Al capped samples as $d_{NiFe}$ decreases. It suggests that the spins reaching at Cu/SiO$_2$ interface are neither absorbed by strong interfacial/bulk SOC nor mostly propagate through spin transparent interface without any backflow. Since SiO$_2$ is an insulating material with negligible spin transparency,[40,41] leading to spin current reflection and spin scattering process which enhances the spin accumulation inside NiFe compared to that of the Al- and Pt-capped sample cases, where the backflow spin currents can further contribute to SOT.[20] As a result, $β_{SOF}$ of SiO$_2$ samples exhibits merely a weak decay as the NiFe thickness decreases [Fig. 4(c)]. These comparative results further demonstrate that for NiFe thicknesses below ~1.7 nm, the spin currents responsible for FL-SOT reach the top Cu/Capping interface and exhibit different interface-dependent spin transport characteristics, corroborating the extended spin dephasing length of FL-SOT in NiFe and highlight the crucial role of interface engineering in tuning spin transport efficiency.[42,43]

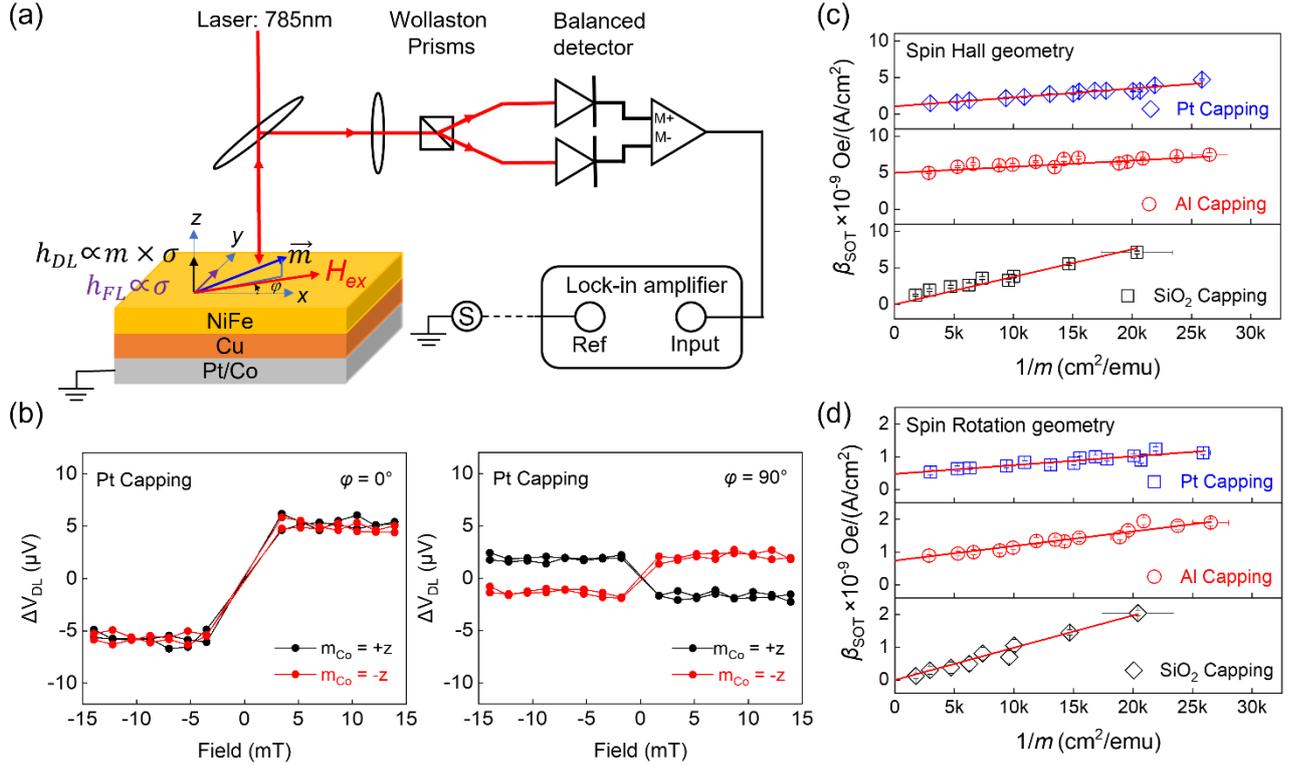

**FIG. 3.** (a) Schematic of DL-SOT detection using polar MOKE. An AC current $I_{ac}$ generates in-plane ($h_{FL}$) and out-of-plane ($h_{DL}$) SOT fields, which tilt the NiFe magnetization. The Kerr rotation is converted to a voltage and detected with a lock-in amplifier. (b) Polar-MOKE signals under Spin Hall geometry (left) and Spin Rotation Geometry (right). (c), (d) DL-SOT efficiency $\beta_{SOT}$ versus $1/m$ for Pt-, Al-, and SiO$_2$-capped samples, obtained in Spin Hall geometry (c) and Spin Rotation Geometry (d).

The DL-SOT effective field ($h_{SH}^{DL}$ and $h_{SR}^{DL}$) was quantified by using a polar magneto-optical Kerr effect (MOKE) setup, as shown in Fig. 3(a), which enables direct detection of out-of-plane magnetization dynamics without complications from planar Hall, anomalous Hall, or thermal effects.[15,25,26,44] An AC current of 50 mA at $f = 70.33$ kHz were applied to suppress $1/f$ noise and enhance the signal-to-noise ratio.[45] The current induced DL-SOT slightly tilts the NiFe magnetization $m_{\text{NiFe}}$ out of plane, producing a modulation of the reflected polarization proportional to $h_{SH}^{DL}(h_{SR}^{DL})$. The MOKE signal $\Delta V_{\text{DL}}$ serves as a direct measure of the effective field $h_{SH}^{DL}(h_{SR}^{DL})$, as expressed in Eq. (5):[44,46]

$$\Delta V_{\text{DL}} = \alpha_{\text{Polar}} \Delta \theta_{h_{\text{DL}}} = \frac{\Delta V(+H_{\text{ex}}) - \Delta V(-H_{\text{ex}})}{2} = \frac{h_{SH}^{DL}(h_{SR}^{DL})}{|H_{\text{ex}}| + M_{\text{eff}}} \quad (5)$$

where $\alpha_{\text{Polar}}$ is the coefficient of polar MOKE response, $\Delta\theta_{h_{\text{DL}}}$ is the magnetization tilt angle induced by $h_{SH}^{DL}(h_{SR}^{DL})$, $H_{\text{ex}}$ is the external magnetic field and $M_{\text{eff}} = M_s - H_{\text{anis}\perp}$ accounts for the effective demagnetization. In the spin Hall geometry, $\Delta V_{\text{DL}}$ is independent of the Co magnetization ($m_{Co}$), as shown in Fig. 3(b) (Left). In contrast, under the spin Rotation geometry [Fig.3(b) (Right)], the MOKE signal reverses polarity upon switching $m_{Co}$, it is consistent with the $h_{SR}^{DL}$ generated through the SRE which is described in Eq. (1). In the calibration procedure, the MOKE signal $\Delta V_{\text{Cal}}$ induced by the calibration field ($h_{Cal}$) can be expressed as:

$$\Delta V_{\text{Cal}} = \alpha_{\text{Polar}}\Delta\theta_{h_{\text{cal}}} = \frac{h_{\text{Cal}}}{|H_{\text{ex}}|+M_{\text{eff}}} \quad (6)$$

where $h_{Cal}$ is the magnetic field amplitude. Therefore, the DL-SOT effective field could be determined by $h_{SH}^{DL}(h_{SR}^{DL}) = \frac{\Delta V_{\text{DL}}}{\Delta V_{\text{Cal}}} h_{\text{Cal}}$. Similar to the data analysis of $\beta_{SOF}$, we are also able to extracted DL-SOT efficiency $\beta_{SOT}$, which is as shown in Fig. 3.

As shown in Fig. 3(c) and Fig. 3(d), the DL-SOT efficiency ($\beta_{SOT}$) scales nearly linearly with $1/m$, consistent with rapid spin absorption at the bottom Cu/NiFe interface and in good agreement with the spin drift–diffusion model.[10,47] A finite $\beta_{\text{SOT}}$ when $1/m$ approaches zero in both Pt- and Al-capped samples, indicating additional interfacial spin-current contributions at the Cu/Pt and Cu/Al interfaces.[8,48–52] However, SiO$_2$-capped stacks show negligible interfacial contributions at the Cu/SiO$_2$ interface. In the spin Hall geometry, interfacial SOC at the Cu/Pt and Cu/Al generates additional conventionally polarized spin currents that directly exert DL-SOT on NiFe.[8,48–50] In contrast, under the spin Rotation geometry, spin-memory loss causes partial depolarization of the spin current,[53–56] producing unconventional spin components that contribute more weakly to the DL-SOT. Consequently, the intercepts obtained in the spin Hall geometry are systematically larger than those in the spin Rotation geometry. Moreover, the larger intercept observed in Al-capped samples compared to Pt-capped ones can be attributed to the weaker interfacial SOC and higher spin transparency of Cu/Al interface,[36,37] which suppress spin absorption and preserve interfacial spin currents, whereas strong SOC in Pt promotes rapid spin dephasing and absorption.[38,39]

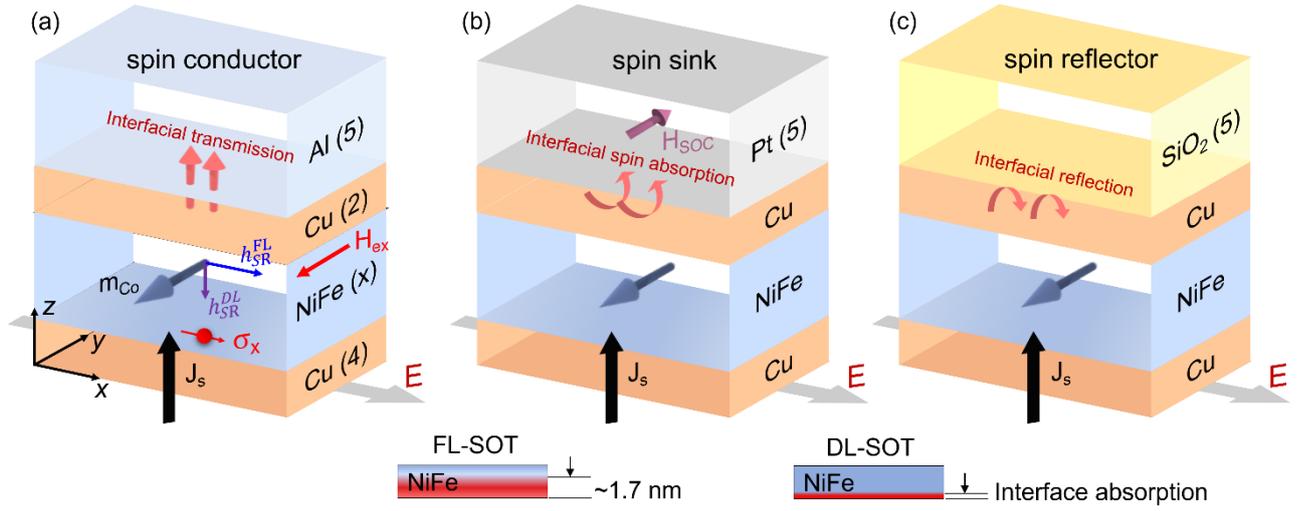

**FIG. 4.** Comparison of FL-SOT propagation in samples capped with (a) Al, (b) Pt, and (c) SiO$_2$. (a) The Cu/Al interface is spin-transparent, permitting efficient spin-current transmission due to its weak SOC; (b) The strong SOC and spin-sink feature of Pt cause substantial spin absorption at the Cu/Pt interface. (c) SiO$_2$ serves as a spin reflector, inducing reflection and scattering at the Cu/SiO$_2$ interface. The backflow spin currents can further contribute to SOT. Notably, the spin currents responsible for FL-SOT exhibit a significantly longer spin dephasing length (~1.7 nm), whereas the DL-SOT is rapidly absorbed at the bottom Cu/NiFe interface.

In summary, we employ an unconventional spin polarization generated by the spin rotation geometry in perpendicularly magnetized Pt/Co/Cu layers to investigate the origin and propagation process of FL-SOT and DL-SOT in NiFe. This geometry eliminates Oersted field interference and enables clear separation of FL-SOT and DL-SOT. To distinguish bulk and interfacial contributions, we introduce a sample-area-normalized moment that accounts for thickness-dependent magnetization and dead layer at the Cu/NiFe interface. DL-SOT follows nearly linear 1/$m$ scaling, consistent with rapid spin absorption at the bottom Cu/NiFe interface, while a finite DL-SOT efficiency $\beta_{SOT}$ when 1/$m$ approaches zero in both Pt- and Al-capped samples reveals additional interfacial spin current contributions at the Cu/Pt and Cu/Al interfaces. In contrast, SiO$_2$-capped samples show negligible interfacial contributions. Moreover, FL-SOT deviates markedly from 1/$m$ scaling and displays a significantly longer spin dephasing length (~1.7 nm) than DL-SOT, indicating extended propagation across the NiFe layer.

Comparative studies further confirm that for NiFe thicknesses below ~1.7 nm, spin currents responsible for FL-SOT reach the top Cu/Capping interface and exhibit different interface-dependent spin transport characteristics, corroborating the extended spin dephasing length of FL-SOT in NiFe. This work provides a unified framework for distinguishing the origin and distinct propagation characteristics of FL-SOT and DL-SOT, offering guidelines for engineering interface-driven SOT efficiency in developing energy-efficient spintronic devices.

## Acknowledgments

This work was supported by the National Natural Science Foundation of China (NSFC) (Grant No. 12104171 and No. 12574119).